\documentclass{aastex}
\usepackage{spr-astr-addons}
\usepackage{url}
\RequirePackage{color}

\newcommand{\emaila}{a78b@yandex.ru}

\begin{document}

\title{Light equation in eclipsing binary CV Boo: \\ third body candidate in elliptical orbit}
\slugcomment{Not to appear in Nonlearned J., 45.}

\shorttitle{Eclipsing binary CV Boo}
\shortauthors{Bogomazov et al.}

\author{A. I. Bogomazov\altaffilmark{1}}
\email{\emaila}
\and
\author{V. S. Kozyreva\altaffilmark{1}}
\and
\author{B. L. Satovskii\altaffilmark{2}}
\and
\author{V. N. Krushevska\altaffilmark{3}}
\and
\author{Yu. G. Kuznyetsova\altaffilmark{3}}
\and
\author{Sh. A. Ehgamberdiev\altaffilmark{4}}
\and
\author{R.G. Karimov\altaffilmark{4}}
\and
\author{A.V. Khalikova\altaffilmark{4}}
\and
\author{M. A. Ibrahimov\altaffilmark{5}}
\and
\author{T. R. Irsmambetova\altaffilmark{1}}
\and
\author{A. V. Tutukov\altaffilmark{5}}
\and

\altaffiltext{1}{M. V. Lomonosov Moscow State University, P. K. Sternberg Astronomical Institute, 13, Universitetskij prospect, Moscow, 119991, Russia}
\altaffiltext{2}{AstroTel Ltd., 1A, Nizhegorodskaya ulitsa, Moscow, 109147, Russia}
\altaffiltext{3}{Main Astronomical Observatory, National Academy of Sciences of Ukraine, 27, Akademika Zabolotnoho ulitsa, Kyiv, 03680, Ukraine}
\altaffiltext{4}{Ulugh Beg Astronomical Institute, Uzbek Academy of Sciences, 33, Astronomicheskaya ulitsa, Tashkent, 100052, Uzbekistan}
\altaffiltext{5}{Institute of astronomy, Russian Academy of Sciences, 48, Pyatnitskaya ulitsa, Moscow, 119017, Russia}

\begin{abstract}
A short period eclipsing binary star CV Boo is tested for the possible existence of additional bodies in the system with a help of the light equation method. We use data on the moments of minima from the literature as well as from our observations during 2014 May--July. A variation of the CV Boo's orbital period is found with a period of $\approx 75$ d. This variation can be explained by the influence of a third star with a mass 
of $\approx 0.4M_{\odot}$ in an eccentric orbit with $e\approx 0.9$. A possibility that the orbital period changes on long time scales is discussed. The suggested tertiary companion is near the chaotic zone around the central binary, so CV Boo represents an interesting example to test its dynamical evolution. A list of 14 minima moments of the binary obtained from our observations is presented.
\end{abstract}


\section{Introduction}

\begin{figure*}[t!]
\centering
\includegraphics[width=0.8\textwidth]{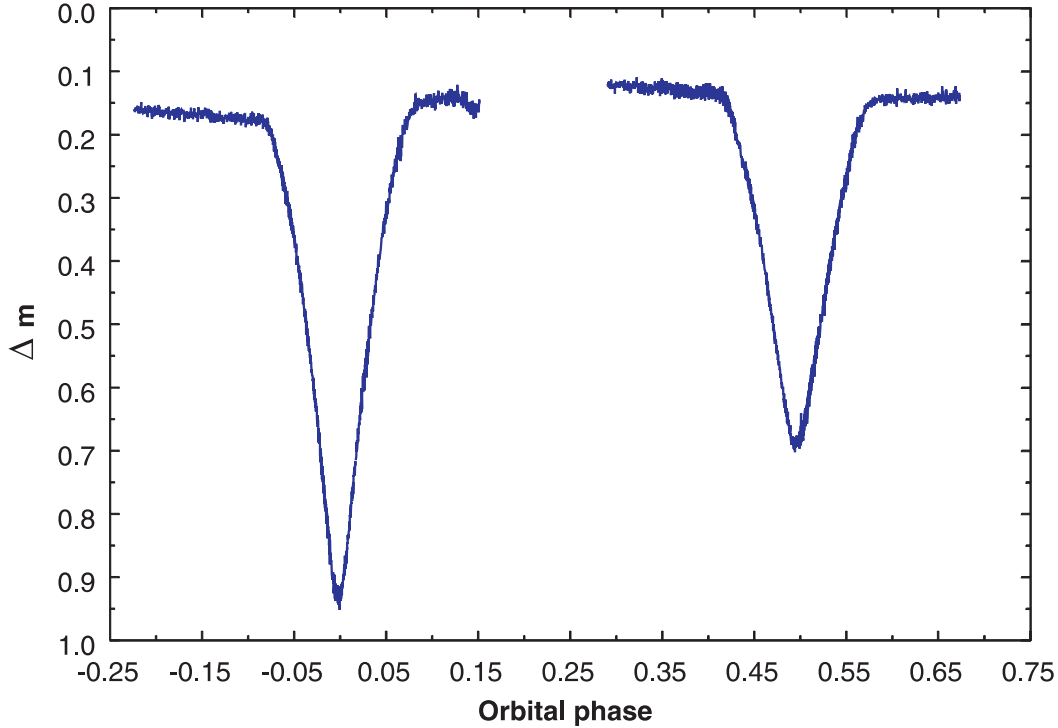}
\vspace{0pt}
\caption{A sample light curve of CV Boo in Bessell {\it R} filter
during nights HJD 2456800, 2456833, 2456858, 2456864.}
\label{fig1}
\end{figure*}

Modern astronomy pays close attention to the physics and evolution of triple and multiple stars. The stability of orbits in such systems, see, e.g. \citep{holman1999,zhuchkov2010}, the influence of additional bodies on the central binary, and the dynamical evolution of orbits in multiple systems, see, e.g. \citep{saito2013,naoz2013,li2014,li2015} are particularly interesting. Some important astrophysical phenomena can be associated with the evolution and dynamical interactions of triple stars. Type Ia supernovae may originate in triples due to the evolution of the central binary under the influence of a third star \citep{iben1999,hamers2013} or due to the dynamical interaction and subsequent direct collision of white dwarfs in triple systems \citep{kushnir2013}. The giant eruption of Eta Carinae in 19th century could be the result of a merger of a massive close binary, triggered by the gravitational interaction with a massive tertiary companion, which is the current member of the binary \citep{pz2016}. Investigations of multiplicity of stars are important for understanding of star formation processes, see, e.g., \citep{tokovinin1999}. This 
makes searches for multiplicity in binary systems of high importance.

According to the General Catalogue of Variable Stars \citep{samus2014}, the orbital period of CV\,Boo is $P=0.8469938$ d and the spectral type of its primary component is G0. CV\,Boo was found as an eclipsing binary by \citet{peniche1985}, and it was studied by many observers during several decades using photographic, visual, PMT (photomultiplier tube), and CCD (charge-coupled device) methods. In a detailed study of CV Boo, \citet{torres2008} presented the results of $V$-band photometric studies and spectroscopic radial velocity (RV) measurements of CV Boo. Using their own data and information from literature, \citet{torres2008} determined the masses, radii and effective temperatures of the primary and secondary components to be $M_1 = 1.032 \pm 0.013$\,M$_{\odot}$, $R_1 = 1.262 \pm 0.023$\,R$_{\odot}$ and $5760 \pm 150$\,K, and $M_2 = 0.968 \pm 0.012$\,M$_{\odot}$, $R_2 = 1.173 \pm 0.023$\,R$_{\odot}$ and $5670 \pm 150$\,K, respectively. \citet{torres2008} found that CV Boo has a light curve variability with a period shorter than the orbital period. They connected this variability with the axial rotation of a spotted component (or both components) which rotates faster than the synchronous rate. \citet{torres2008} noted that this fact disagrees with predictions from the tidal evolution theory according to which both companions should be synchronized. Also, they came to the conclusion that both components are near the end of the main sequence stage and that the primary component even probably entered into the shell hydrogen burning stage. \citet{torres2008} estimated the age of the system as $\approx 9$ Myr, and concluded that the radii of the companion stars in CV Boo are about $10\%$ larger than what is expected from stellar evolutionary models. \citet{torres2008} also claimed that CV Boo has a stable orbital period, based on available 98 primary and 50 secondary minima. Nevertheless, the existence of a large number of high precision CCD photometry obtained since 2008 allows us
to search for an additional companion in CV Boo.

\section{Observations and data reduction}

\begin{table}
\caption{Moments of light minima for CV Boo obtained from our observations. Here ``HJD'' is the Heliocentric Julian Date, ``Min'' is the minimum type (``I'' is the primary minimum, ``II'' is the secondary minimum), $(O-C)_1$ is the difference between observed values of moments of minima and values calculated using Ephemeris (\ref{eph}), $(O-C)_2$ is the difference between observed values of moments of minima and values calculated using Equation (\ref{ephlong}). }
\begin{tabular}{@{}cccc@{}}
\tableline
HJD-2400000 & Min & $(O-C)_1$ & $(O-C)_2$ \\
\tableline
56780.4196 & I & 0.0009 & -0.0009 \\ 
56800.3239 & II & 0.0009 & -0.0010 \\ 
56801.1710 & II & 0.0010 & -0.0009 \\ 
56803.2890 & I & 0.0015 & -0.0003 \\ 
56817.2642 & II & 0.0013 & -0.0005 \\ 
56831.2406 & I & 0.0023 & 0.0005 \\ 
56833.3585 & II & 0.0027 &  0.0009 \\ 
56838.4394 & II & 0.0017 & -0.0002 \\ 
56839.2863 & II & 0.0016 & -0.0003 \\ 
56842.2505 & I & 0.0013 & -0.0006 \\ 
56844.3683 & II & 0.0016 & -0.0002 \\ 
56845.2155 & II & 0.0018 & 0.0000 \\ 
56858.3428 & I & 0.0007 & -0.0011 \\ 
56864.2719 & I & 0.0009 & -0.0010 \\ 
\tableline
\label{minima}
\end{tabular}
\end{table}

CV Boo was observed with the 50 cm AMT-1 telescope with the Apogee Alta-U16M 4Kx4K CCD camera at the Maidanak observatory of the Ulugh Beg Astronomical Institute (Uzbek Academy of Sciences) during 41 nights in spring--summer of 2014. We obtained more than 32\,500 images in the Bessell {\it R} filter with exposures ranging from 8 to 20 s, with a typical exposure of 10 s. Continuos monitoring time intervals were 5-7 hours per night. Bias and dark frames with appropriate exposures were made every night before and after observations. Flat field frames were recorded for the twilight sky.

To process the data we use the aperture photometry method with a specific program\footnote{C-Munipack, http://c-munipack.sourceforge.net}. The optimum aperture corresponds to the minimum of the standard deviation for differential magnitudes. We used 2MASS J15270686+3659270 and 
2MASS J15272880+3647225 as the reference stars. The aperture was constant during each night and didn't change significantly from night to night. Precision values for a single exposure were in the range $0.0024^m-0.004^m$ for different nights. Standard dark and flat field corrections were made. Original fits files were converted to text files containing the CV Boo brightness with respect to the reference stars depending on the Heliocentric Julian Date (HJD).
Light curves of CV Boo were created on the basis of these data. Figure \ref{fig1} plots a sample light curve of CV Boo. In order to achieve the highest possible precision of minima moments we use only the best light curves with both branches around the light curve minimum.

\section{Calculations of moments of minima}

According to \cite{torres2008} at least one of the stars in CV Boo can be spotted. Such spots can influence the symmetry of minima branches. We assumed that positions of spots didn't change during our observations. This assumption is valid because we studied the asymmetry of branches for our observational points and found that it is almost constant. Also, we note that asymmetry of the branches is about an order of magnitude less than the double amplitude of a light effect that is found in the present study. The duration of a minimum of a light curve of CV Boo is $\approx 0.2$ of its orbital period. To calculate moments of minima of a light curve we used a part of a minima with a duration 0.05 of the orbital period. It is equal to $40\%$ of the minimum depth and $25\%$ of the minimum duration.

We calculate moments of minima in free searches by solving the light curves using fixed set of geometrical parameters of the system. A specific program designed to calculate orbital elements is used, see \citep{kozyreva2001} for more details, also see \citep{khaliullina1984} for a similar approach.  We adopted the following parameters of the binary system: radii of the components of $R_1=0.2611$ and $R_2=0.2579$ (in units of the semi-major axis of the system), the inclination angle  $i=87.25^{\circ}$, the zero eccentricity of the central binary orbit, the luminosities of the components $L_1=0.5686$ and $L_2=0.4405$ (in units of the system luminosity), and the limb darkening coefficient $x_1 =x_2=0.55$. This set of parameters is similar to parameters obtained by \citet{torres2008}. The list our moments of minima is presented in Table \ref{minima}.

\section{Secular changes of the orbital period}

\begin{figure*}[t!]
\centering
\includegraphics[width=0.8\textwidth]{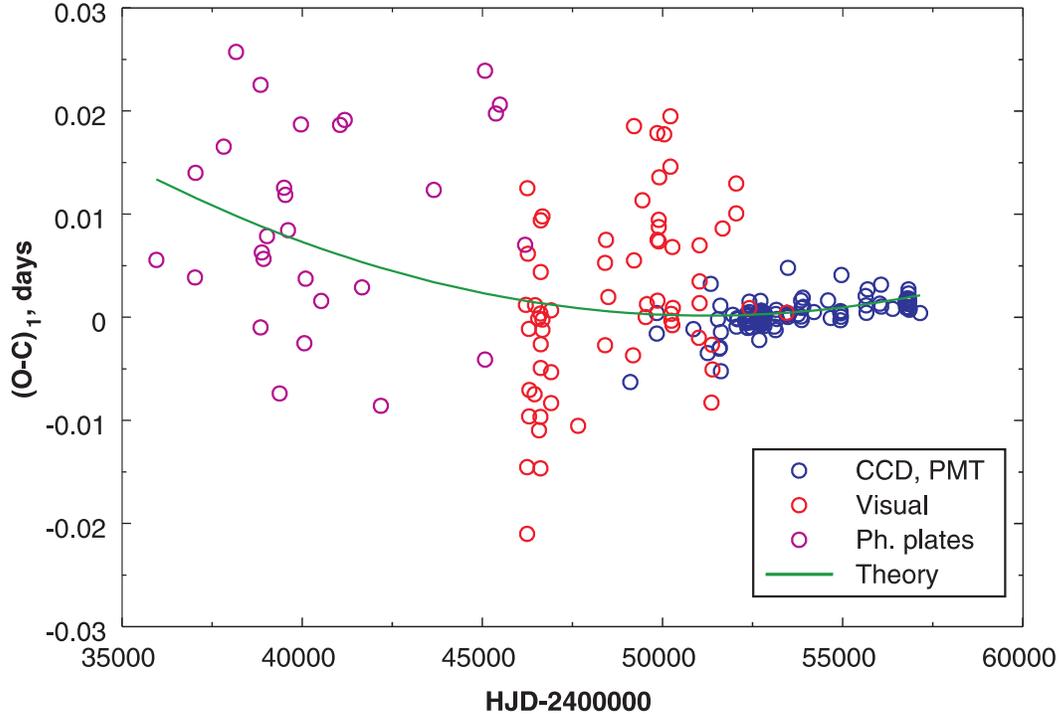}
\vspace{0pt}
\caption{ $(O-C)_1$ diagram for the secular orbital period change of CV Boo. Values $(O-C)_1$ were calculated using all observational moments of minima and Ephemeris (\ref{eph}). Scatter diagrams present observational points for charge-coupled device (CCD) and photomultiplier tube (PMT) observations, for visual observations, and for moments of minima obtained from old photographic plates. A curve presents our theoretical approximation of $(O-C)_1$ using Equation~(\ref{ephlong}) as $O$. }
\label{fig2}
\end{figure*}

As a first step in our study, we used all moments of minima (we collected observational points from \citet{brno}, and use our results from Table \ref{minima}) and the linear Ephemeris obtained by \citet{torres2008}:
\begin{equation}
Min\ I = \text{HJD}\ \ 2452321.845322+0.846993420\times E,
\label{eph}
\end{equation}
\noindent where $E$ is the number of orbital periods since the initial epoch. The moments of the secondary minima are $Min\ II = Min\ I +P/2$, where $P=0.846993420$ d is the orbital period of CV Boo. At this stage, we calculate $(O~-~C)_1$ as the difference between the observational moments of minima and values calculated using Equation (\ref{eph}).

To analyse deviations of moments of minima as a function of time, we used the weighted least squares method because measurements of different authors have different errors. For three groups of observations (``photographic'', ``visual'', and ``PMT/CCD'') we take weights to be equal to the inverse standard deviations in these groups. Polinomial time regressions (with the power in the range from 0 to 3) are checked using two criteria: Student t-criterion (the significance level 0.95), the hypothesis of the zero mean, and $\chi^2$ criterion (the significance level 0.05 for two intervals; only two intervals were used, because observations have different precision), the hypothesis of the equiprobable sign deviations is used. Hypothesises of 0 and 1 powers are declined, for the 3 power the regression matrix is almost degenerate, therefore deviations of moments of minima can be described by a quadratic dependence on time.

A graphical presentation of this result can be found in the Figure \ref{fig2}. As a result we give the next Ephemeris for CV Boo:
\begin{equation}
Min\ I = T_0 + P_0\times E + \frac{\Delta P}{2}\times E^2,
\label{ephlong}
\end{equation}
where $T_0=\text{HJD}\ 2452321.84548$, $P_0=0.846993520$ d, and $\Delta P=8.14\pm 4.07\cdot 10^{-11}$ d.

We estimate the significance of our results using a statistical method by \citet{stellingwerf1978} and calculate the value
\begin{equation}
\label{theta}
\theta=\frac{\sigma^2}{\sigma_0^2},
\end{equation}
where $\sigma_0$ is the standard deviation that corresponds to the values of $(O~-~C)_1$ calculated using Equation (\ref{eph}) and $\sigma$ is the standard deviation that is corrected with the theoretical curve from Equation (\ref{ephlong}). The less $\theta$ means the better coincidence of a theoretical result with observations.

For moments of minima derived from photographic plates, one has $\sigma=0.010$,  $\sigma_0=0.018$, $\theta=0.35$.  For visual moments of minima: $\sigma=0.0089$, $\sigma_0=0.0089$, $\theta=1$. For moments of minima obtained from the CCD and PMT observations: $\sigma=0.0013$, $\sigma_0=0.0012$, $\theta=1.17$. For all points: $\sigma=0.0063$, $\sigma_0=0.0082$, $\theta=0.59$. So, we can conclude that Equation (\ref{ephlong}) describes the whole observations better than the linear ephemeris in Equation (\ref{eph}), therefore the CV Boo orbital period is changing on a long time scale. At the same time, the existing data don't allow us to determine the physical nature of this change.

\section{Short period light equation}

\begin{figure*}[t!]
\centering
\includegraphics[width=0.8\textwidth]{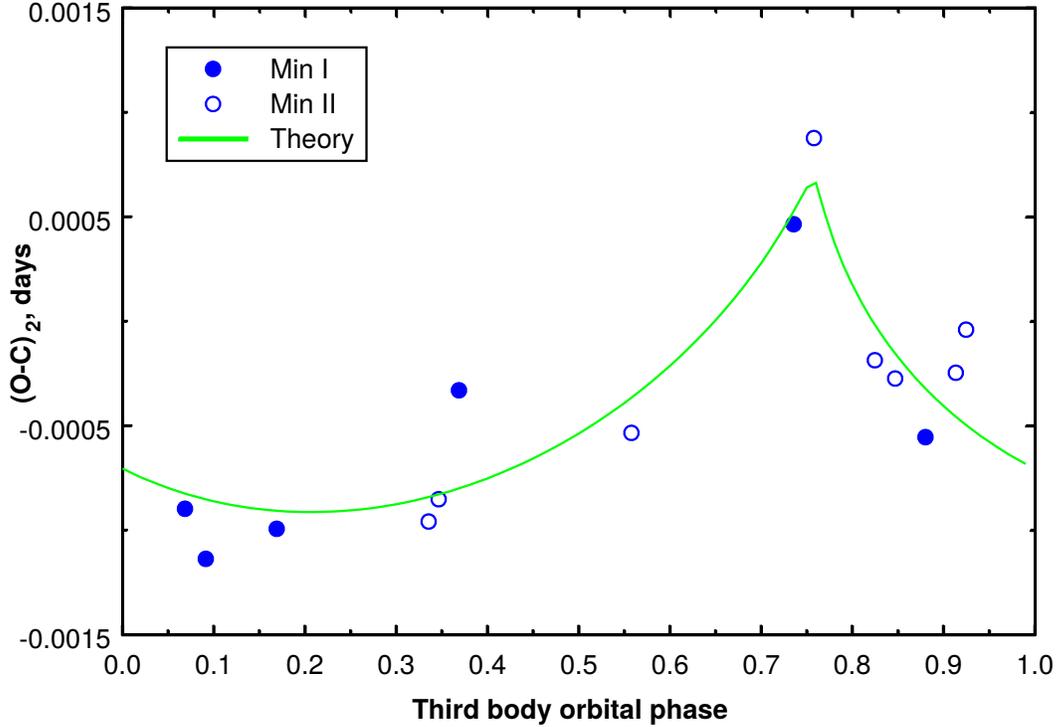}
\vspace{0pt} \caption{$(O-C)_2$ diagram for the orbital period variation of CV Boo, the variation's period is $76.2\pm 1.5$ d. Scatter diagrams present moments of minima derived from our observations in 2014. A curve presents the light time effect (the amplitude is $A=73\pm 10$ s) due to the gravitational influence of a tertiary companion candidate in an eccentic orbit ($e=0.90\pm 0.04$). The initial epoch of the third body is $E_0 = \text{HJD}\  2456775.2\pm 10.0$.}\label{fig3-1}
\end{figure*}

\begin{figure*}[t!]
\centering
\includegraphics[width=0.8\textwidth]{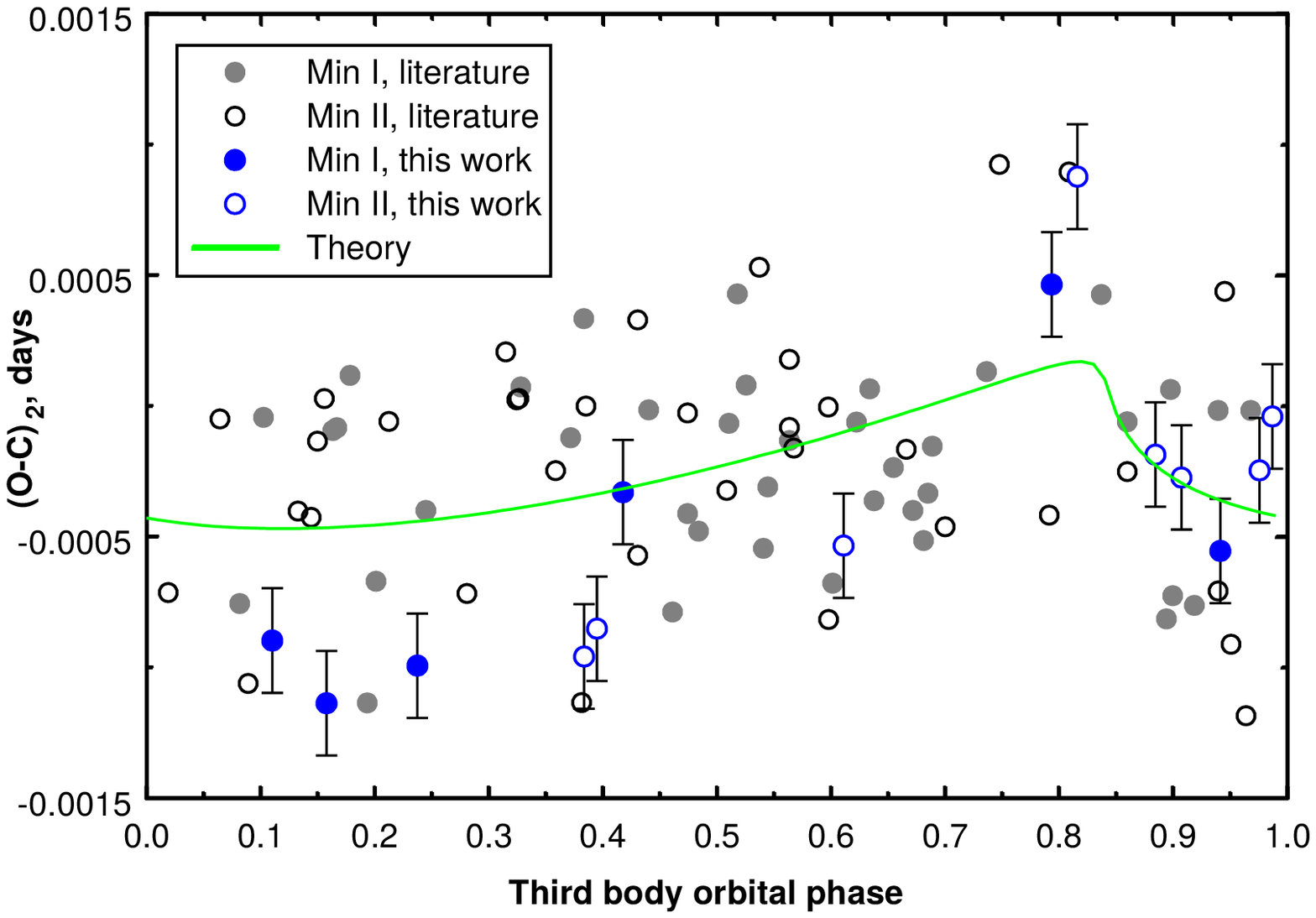}
\vspace{0pt} \caption{$(O-C)_2$ diagram for the orbital period variation of CV Boo, the variation's period is $74.4\pm 0.5$ d. Scatter diagrams present moments of minima derived from our observations and moments of minima taken from literature (after HJD 2452321). A curve presents the light time effect (the amplitude is $48.4\pm 21$ seconds) due to the gravitational influence of a tertiary companion candidate in an eccentic orbit ($e=0.90\pm 0.04$). The initial epoch of the third body is $E_0 = \text{HJD}\  2456772.2\pm 3.5$. Moments of minima are taken form the \citet{brno} and from Table \ref{minima}. }\label{fig3-2}
\end{figure*}

We analyse 14 moments of minima obtained during our observations in 2014 and find that these data probably contain a periodical variation (period $\approx 75$ d) due to the light time effect (see Figure \ref{fig3-1}). Also, we collected published CCD and PMT times of CV Boo light minima since HJD 2452321 in addition to our data. These modern CCD and PMT moments of minima have comparatively low scatter and can be suitable to find a relatively low amplitude variation of the orbital period. We calculate $(O~-~C)_2$ as the difference between the observational moments of minima and the values calculated using Equation (\ref{ephlong}). To derive orbital parameters of a multiple system we solve its light equation (see Equation 3 in \citep{kozyreva2005}, also see \citep{martynov1948} for a more detailed description of this method) by an iterative method of differential corrections.

For our 14 moments of minima from Table \ref{minima} we obtain the following orbital parameters for the third body: the orbital period of $P_3=76.2\pm 1.5$ d, the amplitude of the light equation is $A=73\pm 10$ s, $e_3=0.90\pm 0.04$, $a\cdot \sin i=0.15\pm 0.02$ AU, where $a$ is the semi-major axis of the orbit of the CV Boo binary star's center of mass around the center of mass of the whole system and $i$ is the angle between the third body's orbital plane and the plane of the sky, initial epoch of the third companion is $E_0 = \text{HJD}\  2456775.18\pm 10.0$, the longitude of the pericenter is $\omega_3=112.4^{\circ}\pm 16.0^{\circ}$ for the orbit of the center of mass of the close binary that revolves around the center of mass of the triple system. The result is shown in Figure \ref{fig3-1}, here $\sigma_0=0.000554$, $\sigma=0.000231$, and $\theta=0.17$.

For all CCD and PMT moments of minima after HJD 2452321 taken from the \citet{brno} and from Table \ref{minima} we obtain the following set of light equation parameters: $P_3=74.4\pm 0.5$ d, $A=48.4\pm 21.0$ s, $e_3=0.90\pm 0.04$, $a\cdot \sin i = 0.10\pm 0.04$ AU, $E_0 = \text{HJD}\  2456772.2\pm 3.5$, $\omega_3=156^{\circ}\pm 26^{\circ}$. The result is shown in the Figure \ref{fig3-2}, here $\sigma_0=0.000462$, $\sigma=0.000424$, and $\theta=0.84$.

Among the moments of minima after HJD 2452321 we take 88 moments of minima as ``good'', 16 moments of minima as ``bad''. The reasons to define the moments of minima as ``bad'' and ``good'' points are the following. The moments of minima might be calculated differently by different observers even for the same minimum. Also, the spot activity of the companions can distort the light curves of the system and can lead to differences in calculations of moments of minima. A particular point is assumed to be ``bad'' if its standard deviation exceeds $3\sigma$ value from the short period curve in Figure \ref{fig3-2}. Our 14 points form an uniform group, we used the same method to find all minima moments, light curves with both branches, and spots didn't evolve during our observations, therefore the result based on them is reliable. If we use the linear ephemeris for all points after HJD 2452321 instead of Equation (\ref{ephlong}), the parameters of the light equation slightly differs, but the light equation exists. The difference between $(O-C)_1$ calculated for the theoretical curve in Figure \ref{fig2} for HJD 2452321 and for HJD of the most recent observations (HJD 2456800-2456900) is $\approx 150$ s, the order of magnitude of this value is the same as the order of magnitude of the light equation suggested here. At the same time, we should note that during this period of time the secular orbital period change is practically a smooth function that can be represented as a parabolic function with Equation \ref{ephlong} or as a sinusoidal function with the period $>50$ years. The period of the light effect found in the present study is $\approx 75$ d and so it can be found, because there are about 60 orbital cycles of the suggested tertiary body between HJD 2452321 and HJD 2456900. During our observations the influence of the secular orbital period change is definitely negligible.

The mass function of the suggested tertiary companion is $f(M)\approx 0.02-0.03 M_{\odot}$, see, e. g., \citep{kozyreva2005}, Equation (5). This value yields the lower limit of the mass of this body of $M_3\approx 0.4-0.5 M_{\odot}$.

\section{Conclusions and discussions}

We found a periodical variation of the CV Boo orbital period on a time scale of $\approx 75$ d. This variation can be explained by the light time effect caused by the gravitational influence of a third star with a mass of $\approx 0.4-0.5M_{\odot}$ in an eccentric orbit with $e\approx 0.9$. According to Figure 1 in \citep{shevchenko2014} the third companion is in a chaotic zone or (if the real orbit of the suggested tertiary companion is significantly wider than its projection) almost in a chaotic zone (the values of parameters for the mentioned figure are $\mu=M_2/(M_1+M_2)\approx 0.5$, $r_p\approx
2a_b$, where $M_1$ and $M_2$ are the masses of the primary and the secondary components respectively, $r_p$ is the pericentric distance of the tertiary companion, $a_b$ is the semi-major axis of the central binary's orbit). So the system is very interesting to test numerical calculations of dynamical evolution.

According to \citet{tokovinin2006}, the fraction of solar type spectroscopic binaries with tertiary components reaches 96\% for binaries with the orbital period less than 3 d. Our investigation is in a good agreement with this result. The disagreement between the axial rotation of stars in CV Boo and predictions from tidal evolution theory noted by \citet{torres2008} (at least one of the components rotates faster than the synchronous rate) can probably be explained by a resonance suggested by \citet{borkovits2007}, see the last paragraph of Section 3. This fact serves as an indirect evidence in favour of the existence of a tertiary companion in CV Boo and the influence of this tertiary companion on the central binary.

The physical nature of the secular variation of the CV Boo orbital period can't be clarified with the existing data, so that additional photometric and spectroscopic observations are needed to shed light on this question. It can probably be explained by one of the following mechanisms: (i) the re-synchronization of the axial rotation of one of the companions (or both companions) with orbital rotation of the binary that was disturbed by a third body, (ii) the light time effect caused by the gravitational influence of an additional body with a very long orbital period (about several decades), (iii) the magnetic mechanism \citep{applegate1992,volschow2016} that changes the quadrupole gravitational momentum of the stars in the system by possible strong gravitational disturbances of the orbits in this system, and (iv) the slow mass transfer in the system.

It is possible that RV studies by \citet{popper2000} and by \citet{torres2008} have missed the proposed tertiary companion with the orbital period $\approx 75$ d because of the following reasons. A rough estimate of the orbital velocity of the center of masses of the binary system around the center of masses of the whole system (with the third body's orbital parameters given above) gives $\approx 170-190$ km s$^{-1}$ during the very short periastron passage, the radial velocity is $\approx 130-140$ km s$^{-1}$, during the apastron passage the orbital velocity becomes $\approx 8-10$ km s$^{-1}$, the radial velocity is $\approx 6-8$ km s$^{-1}$. These quantitative estimates are valid for $\omega_3\approx 156^{\circ}$ calculated for 88 points. Orbital velocities of the components of the central binary are about 300 km s$^{-1}$ and their radial velocities range from several tens to more than a hundred of km s$^{-1}$. The orbit of the suggested tertiary companion is an ellipse, therefore most of the time the orbital velocity of the binary should be much less than during the third companion's periastron passage. The radial projection of the orbital velocity can be only less than the orbital velocity and in some positions within the orbit it can be almost zero. Only a few radial velocity measurements were made near the periastron passage of the suggested body. Therefore the presence of this body could be missed in previous spectral studies. An accurate RV study during several months can prove or reject the existence of the candidate tertiary companion found in this work. In case of $\omega_3\approx 112^{\circ}$ for 14 moments of minima obtained from our observations the radial velocities around the periaston and apastron passages are much less than values above, therefore this motion could be missed even easier than in the case of $\omega_3\approx 156^{\circ}$ for 88 moments of minima (our+literature).

A close binary system similar to CV Boo can be a progenitor of a binary system that consists of two carbon-oxygen white dwarfs, therefore it can be a progenitor of a type\,Ia supernova in a triple scenario \citep{kushnir2013}, in this scenario the total mass of colliding white dwarfs can be less than the Chandrasekhar limit. CV Boo itself will most likely merge during further evolution in common envelope stage. To estimate its probable fate we used the web version of the ``Scenario Machine'' \citep{nazin1998}\footnote{See http://xray.sai.msu.ru/sciwork/scenario.html.}. For the masses of the stars equal to the masses of CV Boo components and the semi-major axis in the range $5\div 10R_{\odot}$ and the common envelope efficiency in the range $\alpha_{CE}=0.1\div 10$ we obtain different paths of evolution of this binary. The influence of the tertiary companion can change CV Boo's evolutionary final state: (a) a single white dwarf after merge during common envelope evolution (this is the most probable way of the evolution of this binary), (b) a binary white dwarf, (c) a binary white dwarf, its components collide under the influence of the tertiary companion. The third body is required for the case (c), whereas cases (a) and (b) can be realized both in the presence of the tertiary companion and without it.


\end{document}